\documentclass[aps,prl,twocolumn,showpacs,superscriptaddress]{revtex4}

\usepackage{graphicx}

\begin{document}  

\title{A Laterally Modulated 2D Electron System in the Extreme Quantum
  Limit}

\author{S. Melinte } \affiliation{Department of Electrical
Engineering, Princeton University, Princeton N.J. 08544 }

\author{Mona Berciu } \email[corresponding author:
]{berciu@physics.ubc.ca} \affiliation{Department of Physics and
Astronomy, University of British Columbia, Vancouver B.C. V6T 1Z1,
Canada }

\author{Chenggang Zhou } \affiliation{Department of Electrical
Engineering, Princeton University, Princeton N.J. 08544 }

\author{E. Tutuc } \affiliation{Department of Electrical Engineering,
Princeton University, Princeton N.J. 08544 }

\author{S. J. Papadakis } \affiliation{Department of Electrical
Engineering, Princeton University, Princeton N.J. 08544 }

\author{C. Harrison } \affiliation{Department of Physics, Princeton
University, Princeton N.J. 08544 }

\author{E. P. De Poortere } \affiliation{Department of Electrical
Engineering, Princeton University, Princeton N.J. 08544 }

\author{Mingshaw Wu } \affiliation{Department of Physics, Princeton
University, Princeton N.J. 08544 }

\author{P. M. Chaikin } \affiliation{Department of Physics, Princeton
University, Princeton N.J. 08544 }

\author{M. Shayegan } \affiliation{Department of Electrical
Engineering, Princeton University, Princeton N.J. 08544 }

\author{ R. N. Bhatt} \affiliation{Department of Electrical
Engineering, Princeton University, Princeton N.J. 08544 }

\author{R. A. Register } \affiliation{Department of Chemical
Engineering, Princeton University, Princeton N.J. 08544 }

\date{\today}

\begin{abstract} 
We report on magnetotransport of a two-dimensional electron system
(2DES), located 32 nm below the surface, with a surface superlattice
gate structure of periodicity 39 nm imposing a periodic modulation of
its potential. For low Landau level fillings $\nu$, the diagonal
resistivity displays a rich pattern of fluctuations, even though the
disorder dominates over the periodic modulation.  Theoretical
arguments based on the combined effects of the long-wavelength, strong
disorder and the short-wavelength, weak periodic modulation present in
the 2DES qualitatively explain the data.
\end{abstract}    

\pacs{73.40.Rw, 73.43.Fj, 71.70.Di}

\maketitle


Two-dimensional electron systems (2DESs) subjected to both an
artificial periodic modulation and a quantizing, perpendicular
magnetic field $B$ are expected to exhibit remarkable behavior arising
from the interplay of the period of the modulation and the magnetic
length, and the relative strengths of the periodic modulation,
cyclotron energy ($\hbar \omega_c$), and disorder
potential~\cite{Hofstadter:76,Claro:79, theory}.  When the periodic
modulation's strength is weak compared to $\hbar \omega_c$, each of
the highly degenerate Landau levels (LLs) evolve into an energy
spectrum with recursive properties, the so-called Hofstadter
butterfly. In this limit, the structure of the butterfly is controlled
by the ratio of the magnetic flux through the unit cell area ${\cal
A}$ of the periodic modulation, $\Phi= B{\cal A}$, to the flux quantum
$\Phi_0 =h/e$.  Optimally, $\Phi/\Phi_0$ should be of order unity and
$B$ should be large so that well-separated LLs are formed. This poses
the stringent requirement of a small unit cell area ${\cal
A}$. Additionally, the disorder, present in any real system, should be
small compared to the strength of the periodic potential, so that the
butterfly's larger energy gaps remain open. Then, one may probe
directly the effects of the periodic modulation within a single,
resolved LL. In particular, maxima and minima are expected in the
low-temperature ($T$) diagonal resistivity $\rho_{xx}$ as $B$ is
swept at fixed density $n$ through a LL, i.e. as the Fermi level
($E_F$) moves through the subbands and subgaps of the butterfly. Yet,
for realistic disorder-broadened LLs, new physics can occur, due to
the competition between the strengths of the disorder and the periodic
modulation.

  Here we present low-$T$ magnetotransport measurements in a
GaAs/AlGaAs 2DES, at a distance of 32 nm below the surface, whose
potential is modulated via a surface gate with triangular symmetry and
a periodicity of 39 nm, the smallest periodicity yet
reported~\cite{Schmidt:92,Schlosser:96,Albrecht:01}. Moreover, our
data are taken in the extreme quantum limit, i.e. at low LL fillings
$\nu$ where the integer quantum Hall effect~\cite{Prange:87} is
observed.  Despite the high quality of the 2DES and its close
proximity to the modulating gate~\cite{electrostatics}, the disorder
induced by doping impurities, located at a distance of $\approx$20 nm,
is much larger than the periodic potential.  Naively, one may expect
that in this regime where the periodic potential is only a small
perturbation, it should play no significant role in the transport
properties of the system.  Surprisingly, our results reveal that the
small periodic potential does play a large role.  Experimentally, we
observe that the lowest spin-down LL is significantly broadened and
$\rho_{xx}$ at low $\nu$ exhibits a rich fluctuation pattern.  Our
theoretical arguments, based on the combined effects of the weak,
short-wavelength periodic potential and the strong, long-wavelength
disorder, qualitatively explain the data.  The results highlight the
importance of the very different length scales of the disorder and the
periodic potentials in determining transport through the 2DES: because
there are large areas of the sample where the disorder potential is
very flat, the periodic potential plays the  dominant role locally -- this is one
of the rare cases in physics where a small perturbation has a
significant and non-trivial effect.

  The sample studied here was grown by molecular beam epitaxy, and
consists of a $\rm GaAs/Al_{0.3}Ga_{0.7}As$ heterostructure with a
2DES ($n \approx 3\times 10^{11}\ \rm cm^{-2}$) at a distance $d$ = 32
nm below the surface.  The Si dopant atoms were deposited at a
distance of about 12 nm below the surface.  The low-$T$ mobility,
prior to and after the patterning, is about $3\times 10^{5}\ \rm
cm^{2}/Vs$.  Experiments were performed on 20 $\rm \mu m$-wide Hall
bars fabricated by standard photolithography and wet etching. The
distance between the probes used to measure the resistivity was $20\
\rm \mu m$.  The resistivity coefficients~\cite{rawdata} were measured
by probe pairs located in different regions of the Hall bar, yielding
qualitatively similar results.  The 2DES is placed in a periodic
potential created by a top gate realized by means of a diblock
copolymer nanolithography technique~\cite{Harrison:98}.  First, a
self-assembled layer of hexagonally-ordered polyisoprene nano-domains
(spheres with center-to-center distance $a$) contained in a
polystyrene matrix was formed on the sample's surface. After the
removal of the polyisoprene spheres a polystyrene mask $\approx 15\
\rm nm$-thick is left on the surface and acts as a template for a
Ti/Au metallic gate, whose potential $V_{g}$ can be varied [see right
inset to Fig. 1(a)].  Even without applying an external
gate bias, a rich structure is present in the $\rho_{xx}$ data for
the patterned sample.  This observation corroborates with the results
of previous experiments~\cite{Schlosser:96,Albrecht:01} and
theoretical calculations~\cite{Davies:94} which explain the modulation
at zero bias through the effects of strain and Fermi level pinning at
the semiconductor-metal (gate) interface.

\begin{figure} 
\centering \includegraphics[width=73mm]{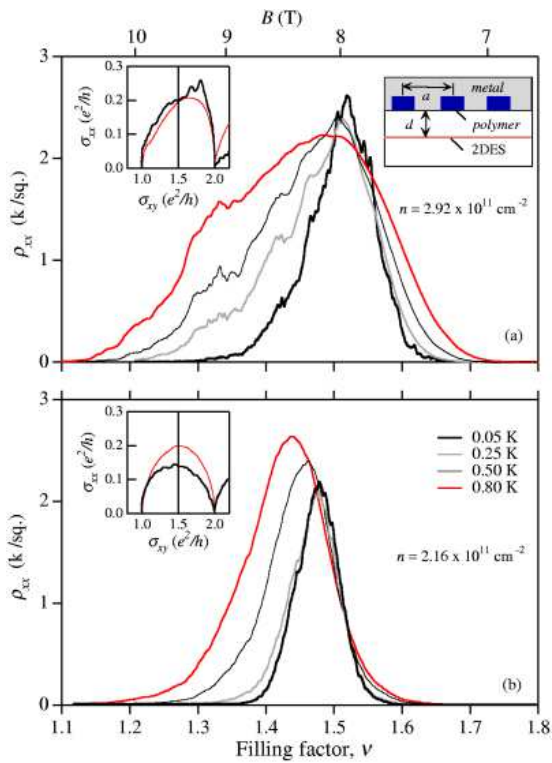}
\caption{ (color online) $\rho_{xx}$ vs $\nu$ traces for (a) the
  patterned sample at $V_{g} = 0 \rm \ V$ and (b) unpatterned sample
  at indicated temperatures. The top magnetic field scale is for the
  data of panel (a).  Left insets: $\sigma_{xx}$ vs $\sigma_{xy}$ at
  $T = 0.05$ and 0.80~K. Right inset: side-view of the device. }
\label{1} 
\end{figure}    

The central result of our experimental work is shown in Fig.~1 where
$\rho_{xx}$ is displayed in the filling factor range $1.1 < \nu <
1.8$ at various temperatures for the patterned and unpatterned
samples~\cite{rawdata}. Several features of the data are noteworthy.
(1) For both samples, $\rho_{xx}$ exhibits a maximum centered around
$\nu=3/2$ and shows reproducible fluctuations at low
$T$~\cite{Jain:88}.  For the modulated sample, however, the observed
fluctuations in $\rho_{xx}$ are much stronger. Qualitatively similar
data was reported in Ref. \onlinecite{Schmidt:92} for a similar device
with 60 nm periodicity, giving further evidence that these
fluctuations are intrinsic. (2) Compared to the data
taken in the unpatterned sample, $\rho_{xx}$ is dramatically
distorted in the patterned sample and has an extended tail on the
low-$\nu$ side.  To better illustrate the observed 
asymmetry for the patterned sample, we display in the insets to Fig.~1
the $\sigma_{xx}$, $\sigma_{xy}$ flow diagram~\cite{Prange:87,rawdata}.  We
note that both the $\rho_{xx}$ fluctuations and  the asymmetry are
reproducible with repeating $B$-sweeps and also 
varying $T$ in the low temperature range.  (3) As $T$ is raised,
concomitantly with the weakening of the $\nu=1$ QHE minimum, new
peaks develop, growing out from the $\rho_{xx}=0$
background.  The peak near $\nu=3/2$ broadens with
increasing temperature and displays at $T= 0.80$~K a full-width at
half-maximum (FWHM) up to four times larger than the low-$T$,
saturated FWHM.  In contrast, the FWHM in the unpatterned sample
increases only by a factor of two for the same $T$ range.  (4) As seen
in Fig.~2, where we present our lowest temperature  $\rho_{xx}$  
as a function of $\nu$ for various gate voltages, the fluctuations
appear predominantly on the low-$\nu$ side of the peak, regardless of
the sign of $V_g$.

\begin{figure}  
\centering \includegraphics[width=71mm]{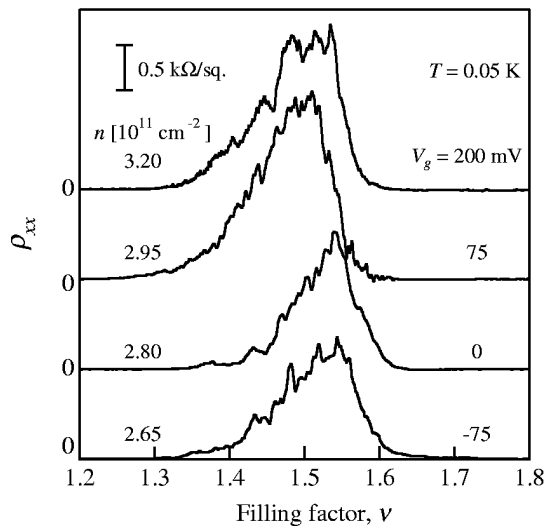}
\caption{ $\rho_{xx}$ vs $\nu$ at $T =0.05$~K and various $V_{g}$
  ($n$).  Curves for $V_{g} = 0$, 75, and 200~mV have been shifted
  for clarity. }
\label{2}  
\end{figure}       

We now discuss the properties of the Hofstadter spectrum for
parameters relevant to our sample. In the absence of disorder, and
neglecting LL mixing, the electronic structure is a function only of
the ratio $\Phi/\Phi_0$.  For $\Phi/\Phi_0=q/p$, with $p$ and $q$
mutually prime integers, the Landau band splits into $q$ subbands.
The Hofstadter butterfly for a repulsive ($V_1>0$) weak triangular
modulation $V(x,y) = 2V_1\left[ \cos{\left( { 2 \pi \over a \sqrt{3}}
(x+\sqrt{3}y)\right)} +\right.$ $\cos{\left( { 2 \pi \over a \sqrt{3}}
(x-\sqrt{3}y)\right)} \ +$ $\left.\cos{\left( { 4 \pi \over a
\sqrt{3}} x\right)}\right]$ is shown in Fig.~3.  The energies of the
eigenstates are referenced to the lowest LL energy $\hbar \omega_c/2$
and are scaled by the amplitude of the first Fourier component of the
periodic potential $V_1$~\cite{Claro:79}.  While the number of total
subbands $q$ jumps discontinuously as $B$ is varied, from Fig.~3 it is
apparent that for all $\Phi/\Phi_0 \in (m, m+1]$, where $m$ is an
integer, the butterfly has precisely $m+1$ main subbands, each of
which has a fractal structure of its own.  Simple arguments show that
the total number of states per unit area in each of the upper $m$ main
subbands is $n_0=1/{\cal A}$, whereas the lowest main subband contains
the rest of $[\Phi/(m\Phi_0) -1] n_0$ states of the LL~\cite{comment}.
The three subgaps shown as yellow regions in Fig.~3 are the most
robust against disorder: if disorder is small, one expects to see
minima in $\rho_{xx}$ when $E_F$ is 
inside these subgaps ($\rho_{xx} \propto \sigma_{xx}$, \cite{rawdata}).  For a 
given $n$ we find the magnetic field $B_i$ at which $E_F$ is
inside each of the main subgaps of the spin-down lowest LL to be $B_i=
{ \Phi_0/2} \left( n + in_0\right)$. For $n=2.92\times
10^{11}$~cm$^{-2}$ ($V_g=0$ in Fig.~1), $E_F$ should cross the main
subgaps at $B_1=7.6$~T, $B_2=9.2$~T, and $B_3=10.7$~T ($\nu=1.59$,
1.31, and 1.13). Figs.~1 and 2 show that the low-$T$
$\rho_{xx}$  vanishes for all $\nu \lesssim1.3$ and $\nu
\gtrsim1.6$, 
suggesting that all states, except for a few in the second highest
subband, are fully localized.  The same conclusion is reached if the
analysis is repeated for an attractive potential $V_1 <0$, for which
the Hofstadter structure of Fig.~3 is inverted.

\begin{figure}  
\centering \includegraphics[width=64mm]{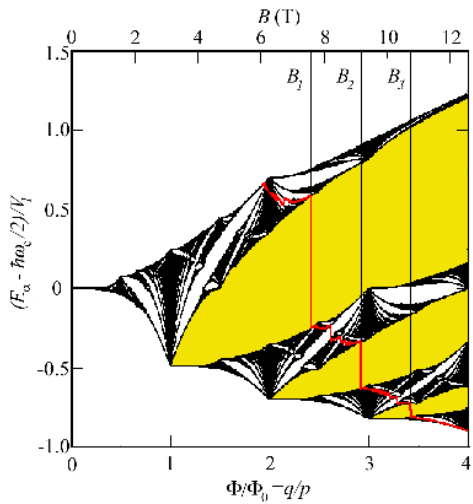}
\caption{ (color) Hofstadter butterfly of the spin-down lowest LL for
  a weak, repulsive, triangular potential modulation. The
  eigenenergies are plotted vs $q/p$, for $p$ and $q$ between 1 and
  150. The position of $E_F$ is shown by the red curve. }
\label{3}    
\end{figure}      

These observations indicate that disorder plays an important role.
 From the measured low-$T$ mobility we estimate the width of the
 disorder-broadened LLs at $0.24 \times (B
 [T])^{1/2}$~meV~\cite{theory}.  For $B=10$~T it is $\approx 0.7$~meV,
 much larger than $V_1 = 0.1 {\ \rm meV/V} \times V_g \rm \ (V)$
 inferred from the exponentially decaying solution of the Laplace
 equation~\cite{electrostatics}.  Although the modulation amplitude is
 not precisely known~\cite{Davies:94}, in our sample it likely remains
 smaller than the disorder.  Since the main subgaps are filled in by
 disorder, the features in $\rho_{xx}$ ($\sigma_{xx}$) cannot be attributed to $E_F$
 crossing the smaller gaps inside the second subband (from $B_1$ to
 $B_2$); these smaller gaps must also be filled in by disorder.
 Hence, an interpretation of our data based on the naive Hofstadter
 butterfly is inappropriate.

\begin{figure} 
\centering \includegraphics[width=82mm]{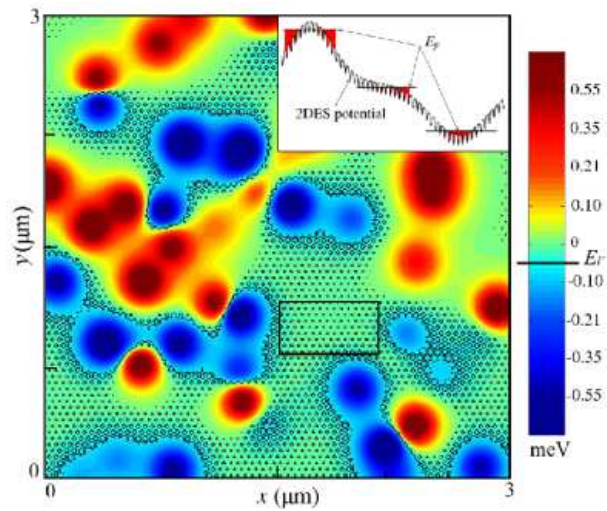}
\caption{(color) The equipotential contours (black curves) of the
total 2DES potential at $\nu=1.3$.  The 2DES potential consists of a
weak, attractive, triangular periodic modulation and a slowly varying
disorder potential. Minima (maxima) of the disorder potential are
shown in blue (red).  The inset schematically shows $E_F$ crossing
through the 2DES potential.}
\label{4}   
\end{figure}   

In order to understand the origin of the $\rho_{xx}$ features, we
must analyze the new regime of a small periodic modulation and large
disorder.  Slowly-varying disorder (expected in high-mobility,
remotely-doped samples such as ours) can be treated in semiclassical
terms, with wave-functions following the equipotentials of the
disorder. The LLs are broadened, with all high (low) energy states
being localized near maxima (minima) of the disorder
potential. Current-carrying extended states correspond to
equipotential contours percolated across the sample and are found in a
so-called critical region at the center of the Landau
band~\cite{Prange:87}. The effect of a supplementary weak periodic
potential is illustrated in Fig.~4, where we plot equipotential
contours of the total 2DES potential. The disorder potential is a sum
of screened Coulomb potentials from doping impurities located 20 nm
from the 2DES~\cite{Zhou}. The amplitude of both disorder and the
periodic potentials are close to experimental estimations.  Instead of
smooth trajectories, the equipotential contours now have a fractured
nature, with many little ``bubbles'' around minima of the periodic
potential on the flat areas of the disorder potential.  Since the
lattice constant $a$ is comparable to the magnetic length, {\em
quantum mechanical tunneling spreads the electronic wave-functions
over these flat regions, and considerably helps their percolation
throughout the sample}.  This picture is in qualitative agreement with
the results of Figs.~1 and 2, which show that $\rho_{xx}$ of the
patterned sample is considerably enhanced with increasing $T$ on the
low-$\nu$ side, suggesting transport through temperature-activated
hopping between nearby states, absent in the unpatterned sample.

This picture also offers a possible explanation for the detailed peak
and valley structure observed in our magnetotransport data. Consider
the area enclosed in the rectangle drawn in Fig.~4, which is almost
disorder-free (it has very little underlying disorder
variation). Wave-functions across such flat regions
correspond to those of finite-area Hofstadter butterflies, with
appropriate boundary conditions.  For the appropriate energy range,
$E_F$ is either inside a subband or a gap of such local Hofstadter
structures. If $E_F$ is inside a subband, wave-functions span the
flat region and help enhance the percolation.  When $E_F$ is inside a
gap, there are no states supported by the flat region and
hence, no current can flow across it. Such
flat regions act as switches which are turned on or off as $E_F$ is
changed, helping or hindering the current flow through the sample.
When one or more switches are off, there are fewer paths for the
current to be carried across the sample, and a valley is expected in
$\sigma_{xx}$ ($\rho_{xx}$).

Finally, the appearance of the fluctuation pattern preponderantly on the
low-$\nu$ side (see Fig.~2) is a direct 
consequence of the particle-hole asymmetry of the triangular
potential.  For LL filling factors $1<\nu<3/2$ ($3/2<\nu <2$),
percolation between wave-functions localized around minima (maxima) of
the disorder is helped by the ``bubbles'' centered on the minima
(maxima) of the periodic potential (see inset to Fig.~4). When
attractive, the triangular potential has deep minima on a triangular
pattern, and shallow maxima on a displaced honey-comb pattern (notice
that there are almost no honey-comb arranged ``bubbles'' in Fig.~4).
For $3/2<\nu <2$, the shallow maxima of the periodic potential are not
as effective in enhancing percolation, and $\sigma_{xx}$ ($\rho_{xx}$) in
the modulated 
sample is similar to that of the unpatterned sample.

\begin{figure}[t] 
\centering \includegraphics[angle=270,width=72mm]{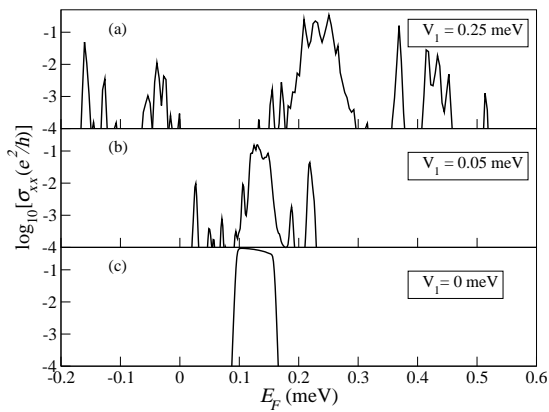}
\caption{ $\sigma_{xx}$ vs $E_F$ for a $3.1\times 2.9 \rm \ \mu m^{2}$
  sample at $T=0.01$~K and indicated $V_1$ for $q/p = 5/2$. The
  standard deviation of the disorder potential is 0.7~meV.  $E_F$ is
  measured from $\hbar\omega_c/2$.  }
\label{5}  
\end{figure}    

These semi-classical arguments are supported by preliminary
computations of $\sigma_{xx}$, shown in Fig.~5, based on the
Kubo-Landauer formalism~\cite{Zhou}. The calculation is done for a
given realization of a slowly-varying disorder potential and for a
fixed value of $B$. As $n$ is varied from 2.6 to $3.4\times
10^{11}$~cm$^{-2}$, $E_F$ sweeps through the disorder broadened LL. In
Fig.~5(a), the peak-to-peak amplitude of the periodic modulation --
$9V_1$ for the triangular superlattice -- is somewhat larger than
disorder's strength.  Well-defined subbands start to separate and
evolve towards the Hofstadter butterfly structure expected in the
limit where disorder becomes negligible compared to the periodic
modulation.  In the opposite limit of no periodic modulation
[Fig.~5(c)], we observe a narrow, smooth peak when $E_F$ is inside the
critical region of percolated states (not centered at $E=0$ because
the disorder potential is not particle-hole symmetric). For a periodic
modulation weaker or comparable to the disorder strength [Fig. 5(b)],
a pattern of peaks and valleys emerges, as the periodic potential
influences the percolation inside and near the critical
region. Efforts to improve the description of the disorder potential
are under way, so that quantitative comparisons with the experiment
become possible.


 To conclude, our work extends the rich problem of modulated 2DESs in
quantizing magnetic fields to a new regime of long-wavelength, strong
disorder and short-wavelength, weak periodic potential, and uncovers
yet more interesting physics.

This work is supported by a NSF/NATO Grant, a NSF MRSEC Grant and by
NSERC of Canada (M.B.).



\begin{thebibliography}{10}
\bibitem{rawdata} The longitudinal and transverse resistivities
  $\rho_{xx}$ and $\rho_{xy}$ were recorded at both positive and 
  negative $B$ to eliminate the mixing between transport tensor
  components. 
$\sigma_{xx}$ and $\sigma_{xy}$ were then determined from  the standard relations
  $\sigma_{xx}=\rho_{xx}/(\rho_{xx}^2+\rho_{xy}^2)$;
  $\sigma_{xy}=\rho_{xy}/(\rho_{xx}^2+\rho_{xy}^2)$.
\end{thebibliography}
\end{document}